\documentclass{article}
\usepackage{spconf,amsmath,graphicx}
\usepackage{float}
\usepackage{booktabs,multirow}

\title{Pre-Trained Model Representations and their Robustness against Noise for Speech Emotion Analysis}
\name{Vikramjit Mitra, Vasudha Kowtha, Hsiang-Yun Sherry Chien, Erdrin Azemi, Carlos Avendano}
\address{Apple}
\begin{document}
%
\maketitle
\begin{abstract}
Pre-trained model representations have demonstrated state-of-the-art performance in speech recognition, natural language processing, and other applications. Speech models, such as Bidirectional Encoder Representations from Transformers (BERT) and Hidden units BERT (HuBERT), have enabled generating lexical and acoustic representations to benefit speech recognition applications. We investigated the use of pre-trained model representations for estimating dimensional emotions, such as activation, valence, and dominance, from speech. We observed that while valence may rely heavily on lexical representations, activation and dominance rely mostly on acoustic information. In this work, we used multi-modal fusion representations from pre-trained models to generate state-of-the-art speech emotion estimation, and we showed a ${100\%}$ and ${30\%}$ relative improvement in concordance correlation coefficient ($CCC$) on valence estimation compared to standard acoustic and lexical baselines. Finally, we investigated the robustness of pre-trained model representations against noise and reverberation degradation and noticed that lexical and acoustic representations are impacted differently. We discovered that lexical representations are more robust to distortions compared to acoustic representations, and demonstrated that knowledge distillation from a multi-modal model helps to improve the noise-robustness of acoustic-based models.
\end{abstract}
\vspace{-1mm}
\vspace{-2mm}
\section{Introduction}
\label{sec:intro}
\vspace{-2mm}
Speech-based emotion recognition models aim to recognize the emotional state of a speaker in real-time. Current human-machine interaction systems can recognize the words said by the speaker but fail to acknowledge the expressed emotion. Reliable and robust speech-based emotion models can help improve human-computer interaction and health/wellness applications, such as voice assistants \cite{mitra2019leveraging, kowtha2020detecting}, clinical mental health diagnoses \cite{sahana2015automatic, stasak2016investigation}, and/or therapy treatments \cite{torre2018putting}.

Emotion recognition research has pursued two definitions of emotion: (1) discrete emotions \cite{ekman1992argument, plutchik2001nature}, and (2) dimensional emotions \cite{russell1977evidence}. Discrete emotions are categorized as fear, anger, joy, sadness, etc., and the emotion lexicon size varies from six \cite{ekman1992argument} to twenty-seven categories \cite{cowen2017self}. Variation in discrete emotion definitions results in annotation complexities, difficulty in realizing consistent emotion labels, and failure to include rare and/or complex emotional states. Dimensional emotion  represents the emotion space using a three-dimensional model of (i) \textit{Activation} (reflects vocal energy), (ii) \textit{Valence} (indicates negativity or positivity), and (iii) \textit{Dominance} (specifies how strong or meek one may sound).

Prior work have shown that combining lexical- and acoustic- based representations can boost model performance on emotion recognition from speech \cite{sahu2019multi, ghriss2022sentiment, srinivasan2021representation, siriwardhana2020jointly}. Recent studies have demonstrated that pre-trained model representations can generate state-of-the-art emotion recognition performance \cite{siriwardhana2020jointly, pepino2021emotion}, which can substantially improve valence estimation performance \cite{srinivasan2021representation, mitra2022Speechemotion}. For real world applications, it is imperative to assess the robustness of speech emotion models against background distortions. The effect of noise in speech emotion recognition and robustness of features and modeling approaches have been studied in \cite{leem2022not, triantafyllopoulos2019towards}, however, prior art have not investigated noise robustness of pre-trained model representations. 

\noindent In this work we demonstrate:
\newline 
(1) Fusion of multiple lexical and acoustic representations obtained from pre-trained models results in \textit{state-of-the-art} dimensional emotion estimation performance on a large publicly available real-world data-set.
\newline
(2) Pre-trained model representations (both lexical and acoustic) demonstrate better robustness against noise and channel degradation compared to low-level acoustic features. \newline
(3) Lexical embeddings are more robust to acoustic distortions than their acoustic counterparts. 
\newline
\vspace{-6mm}

\section{Data}
\vspace{-2mm}
All experiments are run using the publicly available MSP-Podcast dataset 1.6 \cite{mariooryad2014building} containing speech from English speakers. The speech segments contain single speaker utterances with a duration of 3 to 11 seconds. The data contains manually assigned valence, activation and dominance scores (7-point likert scores) from multiple graders. The data split is shown in Table \ref{tab:table1}. To make our results comparable to literature \cite{ghriss2022sentiment, srinivasan2021representation}, we report results on Eval1.3 and Eval1.6 (see Table \ref{tab:table1}). A subset of the data has manual transcriptions which was used to perform the ASR-analyis presented in Section 4.

To analyze the robustness of the emotion models, we add noise and reverberation to the Eval1.6 set. Noise (reflecting home appliance noise) was added at 3 SNR levels: 20 to 30 dB (avg. 20 dB), 10 to 20 dB (avg. 15 dB), and 0 to 10 dB (avg. 5 dB). The SNR level for each file was randomly selected from a uniform distribution. Hence, we created three additional evaluations sets one for each of the three SNR levels (see Table \ref{tab:table1}). The noise-added data was artificially reverberated to simulate realistic indoor conditions. For performing noise-aware training, we added similar distortions to 9\% of the files from the training set "Train+noise" in Table \ref{tab:table1}. 
\vspace{-4mm}
\begin{table}[htb]
\centering
\caption{MSP-podcast data split and noise-degraded sets.}
\vspace{1mm}
\begin{tabular}{lll}
\hline
\textbf{Split} & \textbf{hours} & \textbf{Description}\\
\hline
Train1.6 & 85 & Training set \\
Valid1.6 & 15 & Validation set \\
Eval1.3 & 22 & Podcast1.3 evaluation set \\
Eval1.6 & 25 & Podcast1.6 evaluation set \\
Train+noise & 93 & Training set with 9\% noisy data \\
$Eval1.6_{25dB}$ & 25 & Eval1.6 + noise within 20-30 dB \\
$Eval1.6_{15dB}$ & 25 & Eval1.6 + noise within 10-20 dB \\
$Eval1.6_{5dB}$ & 25 & Eval1.6 + noise within 0-10 dB \\
\hline
\end{tabular}
\label{tab:table1}
\end{table}
\vspace{-5mm}
\section{Representations \& Acoustic Features}
\subsection{Acoustic Features}
\vspace{-2mm}
As a baseline acoustic feature we use 40-dimensional mel-filterbank energies appended with pitch, pitch-delta, and voicing features, resulting in a 43-dimensional $MFBF_0$ feature. 
\vspace{-7mm}

\subsection{Acoustic Representations from HuBERT}
\vspace{-2mm}
We explore embeddings generated from HuBERT large, a pre-trained acoustic model \cite{hsu2021hubert}, which was pre-trained on 60,000 hours of speech from the Libri-light dataset with 24 transformer layers and 1024 embedding dimensions. In our study, we extracted the following embeddings: \\
(1) $HUBERT_L$ embeddings from the $24^{th}$ layer of the pre-trained HuBERT large model (no fine tuning).
\newline
(2) $HUBERT_{A}$ embeddings from the $HUBERT_L$ model ($24^{th}$ layer) fine-tuned on an automatic speech recognition (ASR) task using 100 hours of Librispeech data. 
\vspace{-3mm}

\subsection{Lexical Representations from BERT}
\vspace{-2mm}
We investigated the use of pre-trained BERT \cite{kenton2019bert} embeddings for emotion model training. As ground truth transcriptions are only available for a subset of speech segments in MSP Podcast data, we rely on two ASR models to estimate transcripts from speech: $HUBERT_{A}$ and an in-house ASR system. Given transcribed speech, we obtained 768-dimensional embeddings from the $12^{th}$ layer of the pre-trained BERT model. 
\vspace{-4mm}
\section{Analysis of text-data from ASR models}
\begin{figure*}[!ht]
\begin{minipage}[b]{1.0\linewidth}
  \centering
  \centerline{\includegraphics[width=5.5in]{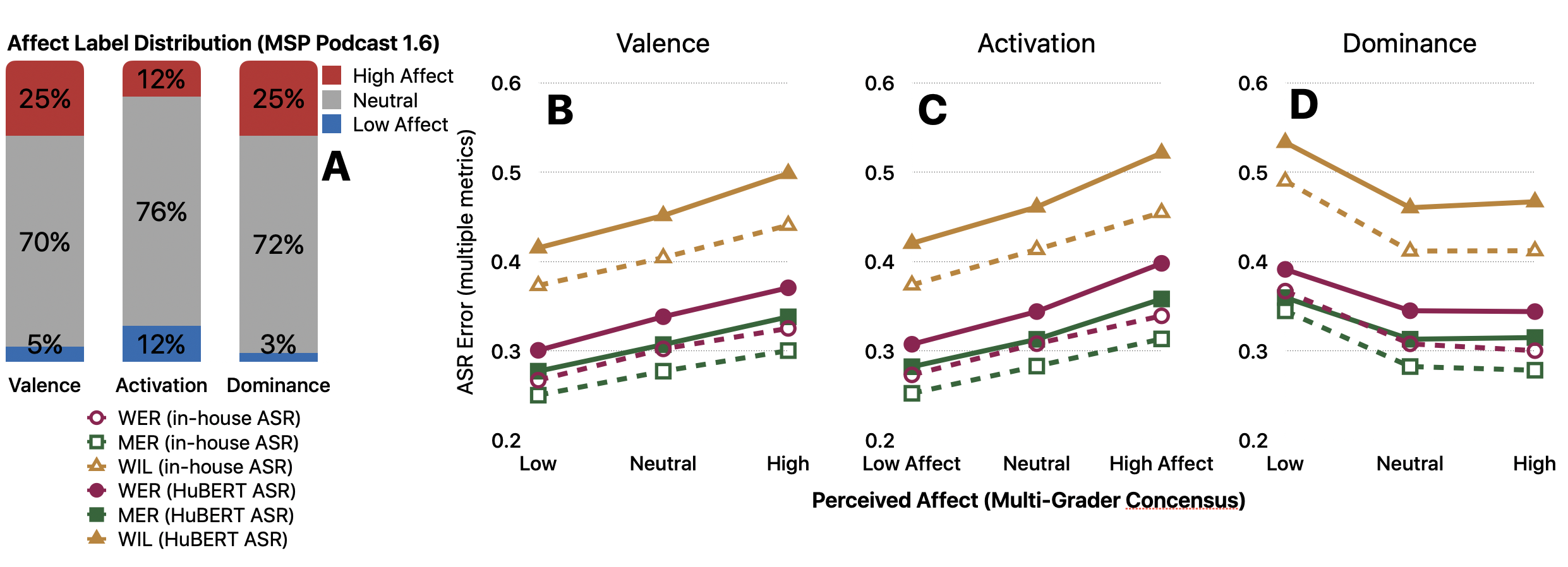}}
\end{minipage}
\caption{WER variation with low, neutral and high values of Valence, Activation and Dominance, obtained from transcriptions generated by the ASR models (WER: word error rate, WIL: word information lost, MER: match error rate)}
\label{fig:fig1}
\end{figure*}

\vspace{-2mm}
We evaluated ASR performance on a portion of the MSP-Podcast data that contained manual transcription. We used two separate ASR systems (an in-house ASR system and the $HUBERT_{A}$ system) and Figure \ref{fig:fig1} shows how the word error rate (WER) varied for different ranges (low, neutral and high) of valence, activation, and dominance by treating each dimension as independent of the others. Figure \ref{fig:fig1} show that for higher values of valence and activation the WER increases, while the same is observed at lower dominance. The increase in WER at high valence, high activation or low dominance indicate that ASR systems made more mistakes under such conditions. Figure \ref{fig:fig1} indicates that BERT embeddings from ASR transcriptions may be impacted by varying WERs at different emotion levels. 
\vspace{-2mm}

\section{Emotion Estimation Model}
\vspace{-2mm}
\subsection{Model Architectures}
\begin{figure}[htb]
\begin{minipage}[b]{1.0\linewidth}
  \centering
  \centerline{\includegraphics[width=4.3cm]{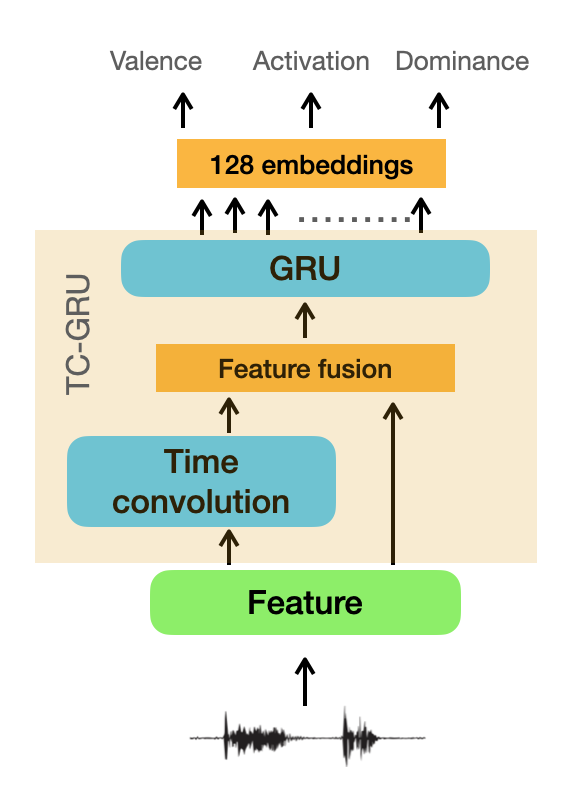}}
\end{minipage}
\caption{The basic TC-GRU network }
\label{fig:fig2}
\end{figure}

\begin{figure}[htb]
\begin{minipage}[b]{1.0\linewidth}
  \centering
  \centerline{\includegraphics[width=3.4in]{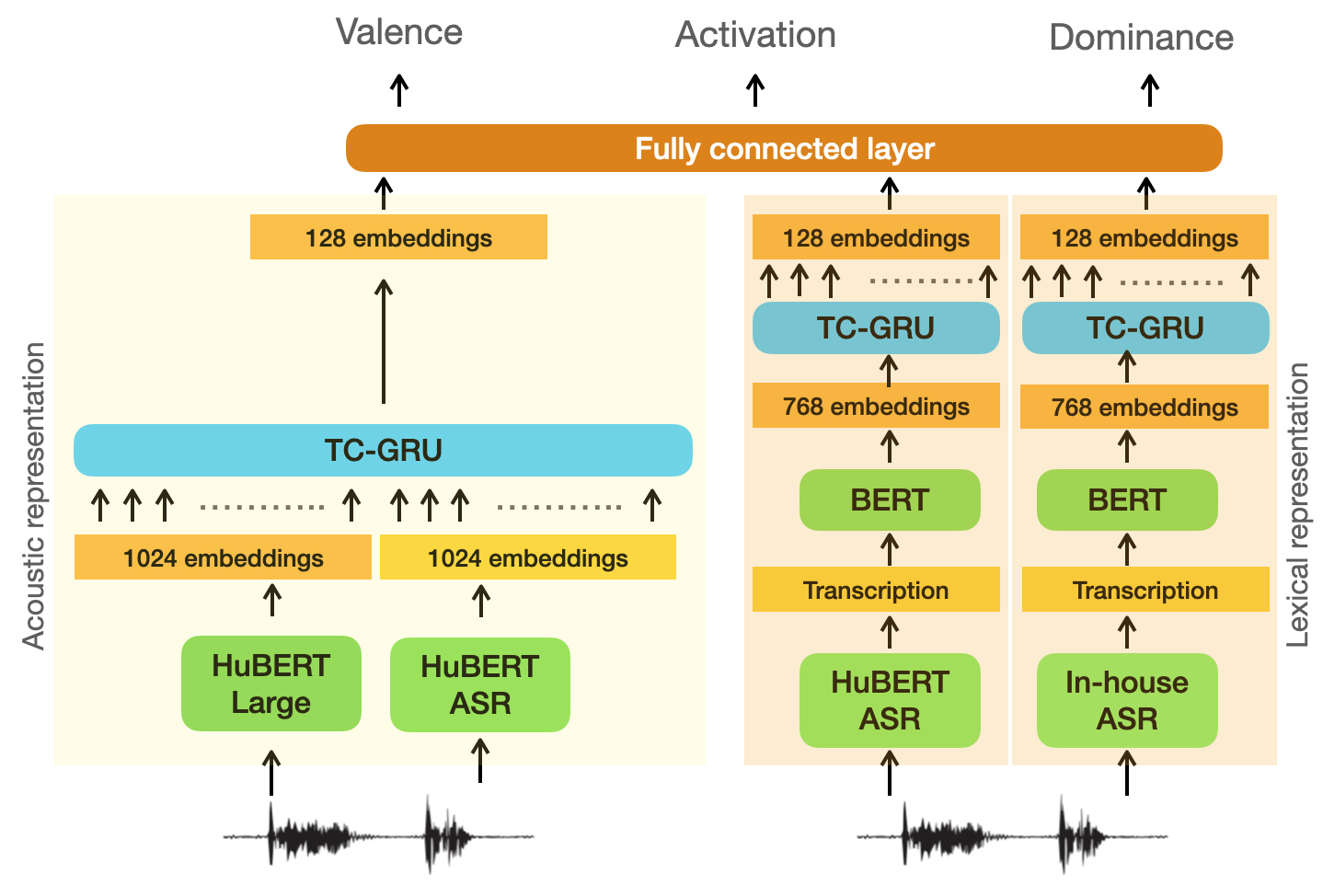}}
\end{minipage}
\caption{Multi-modal emotion estimation model}
\setlength{\belowcaptionskip}{0pt}
\label{fig:fig3}
\end{figure}

\vspace{-2mm}
All models presented here use the same basic architecture (see Figure \ref{fig:fig2}) for supervised dimensional emotion estimation: a time-convolutional (TC) layer with skip connection followed by 2-layer GRUs whose output is projected to a 128-dimensional embedding layer, followed by an output layer. 
\newline
\textit{Uni-Modal Input}: The TC-GRU model takes one modality as input (any one of the features described in Section 3). The input feature dimension and time resolution may vary based on the feature type and its modality.
\newline
\textit{Multi-Modal Input}: The lexical-acoustic fusion model uses a combination of $HuBERT_L$ and $HuBERT_A$ embeddings (2048 dimension) which are projected to 128 dimension as shown in Figure \ref{fig:fig3}. The resulting 128 dimensional embeddings are concatenated with embeddings obtained from TC-GRU models trained with BERT embeddings from $HuBERT_A$ and in-house ASR transcriptions, respectively. 
\vspace{-1mm}
\vspace{-1mm}

\subsection{Model Training}
\vspace{-2mm}
For uni- and multi- modal inputs, 2-layer TC-GRU models were trained, where the performance on a held-out Valid1.6 set (see Table \ref{tab:table1}) was used for model selection. Concordance correlation coefficient ($CCC$) \cite{lawrence1989concordance} is used as the loss function ($L_{ccc}$) (see (\ref{eq1})), where ${L_{ccc}}$ is a combination ($\alpha=1/3$ and $\beta=1/3$) of $CCC$'s obtained from each of the three dimensional emotions. $CCC$ is defined by (\ref{eq2}), where ${\mu _{x}}$ and ${\mu _{y}}$ are the means, ${\sigma _{x}^{2}}$ and ${\sigma _{y}^{2}}$ are the corresponding variances for the estimated and ground-truth variables, and ${\rho}$ is the correlation coefficient between those variables. The models are trained with a mini-batch size of 32 and a learning rate of 0.0005. 
\vspace{-2mm}
\begin{multline}
  L_{ccc}= - (\alpha CCC_{v}+\beta CCC_{a}+(1-\alpha-\beta)CCC_{d})
  \label{eq1}
\end{multline}
\begin{equation}
    CCC = \frac {2\rho \sigma_x\sigma_y}{\sigma_x^2+\sigma_y^2 +(\mu_x-\mu_y)^2 }.
    \label{eq2}
\end{equation}
\vspace{-5mm}

\subsection{Distillation}
\vspace{-2mm}
To investigate if knowledge can be distilled from the multi-modal models into the simpler uni-modal models, we explored distilling the representations learned by the former into the latter using a teacher-student framework. Let the embedding from the multi-modal teacher network $T$ for a speech sample ${i}$ is ${E_{T,i}}$, the learned embedding from the student model $S$ for sample $i$ be ${\hat{E}_{S,i}}$, the ground truth label be ${l_{k,i}}$ (where ${k\in \{a,v,d\}}$), and the estimated label from $T$ are ${\tilde{l}_{k,i}}$. We propose the distillation loss as:
\vspace{-1mm}
\begin{equation}
  L_{dis,k} = CCC_{E_{T,i},\hat{E}_{S,i}} \cdot \gamma_{i},
  \label{eq3}
\end{equation}
\begin{equation}
  \gamma_{i} = 1 - \frac {1}{M} \left ( \frac {\sum_k |l_{k,i} - \tilde{l}_{k,i}|}{3} \right ).
  \label{eq4}
\end{equation}

In (\ref{eq3}) the distillation loss is computed using the $CCC$ (see \ref{eq2}) between the teacher (${E_{T,i}}$) and the student (${\hat{E}_{S,i}}$) embeddings, and is weighted by the residual error ($|l_{k,i} - \tilde{l}_{k,i}|$) between the target (${l_{k,i}}$) and the estimated (${\tilde{l}_{k,i}}$) labels of the teacher network. We explored $L_2$ and $cosine$ distance loss for distillation and found the proposed $CCC$ based loss to perform better. Note that $M$ represents the dynamic range of the target labels ${l_{k,i}}$ in the dataset. $\gamma_{i}$ provides a confidence measure for each sample, where it is closer to 1 for samples that have low residual error from the teacher network and closer to 0 when the residuals are high. The final loss is defined as: 
\vspace{-2mm}
\begin{equation}
  L_{k} = \kappa (L_{ccc,k} + L_{CE,k}) + \lambda L_{dis,k},
  \label{eq5}
\end{equation}
where the values of $\kappa$ and $\lambda$ were empirically selected to be 0.001 and 1, to emphasize more on the distillation loss.
\vspace{-1mm}

\section{Results}
\vspace{-2mm}
\subsection{Emotion estimation}
\vspace{-2mm}
We trained unimodal TC-GRU emotion models with the following inputs: (i) ${MFBF_0}$ based acoustic features, embeddings extracted from (ii) $HuBERT_L$ and (iii) $HuBERT_A$, and (iv) lexical embeddings from BERT ($BERT_{HuB}$) using transcriptions generated by $HuBERT_{A}$ (note that BERT embeddings from in-house ASR demonstrated similar performance as $BERT_{HuB}$). We trained the multi-modal model (Figure \ref{fig:fig3}) using speech embeddings ($HuBERT_L$ and $HuBERT_{A}$ as input, combined with the $BERT$ embeddings. The performance of the uni- and multi- modal models for dimensional emotion estimation is shown in Table \ref{tab:table2}. Table \ref{tab:table2} shows that features extracted from any given choice of pre-trained model (whether lexical or acoustic), significantly improves valence estimation relative to ${MFBF_0}$. $BERT$ and $HuBERT_L$ based models demonstrated a $70\%$ and $60\%$ relative improvement in $CCC$ on valence estimation compared to ${MFBF_0}$, respectively. Multi-modal fusion further improved valence estimation relative to all uni-modal models (see Table \ref{tab:table2}), where we observe a $100\%$, $40\%$ and $44\%$ relative improvement in $CCC$ on valence estimation compared to the ${MFBF_0}$, $BERT$ and $HuBERT_L$ based models, respectively. Table \ref{tab:table2} shows that lexical representations may not be sufficient by themselves for estimating activation and dominance, however, through multi-modal fusion they achieve the best overall performance. 
\vspace{-2mm}
\subsection{Model Robustness to Acoustic Distortions}
\vspace{-2mm}
We investigated the robustness of the models presented above, to acoustic distortions unseen during training. Table \ref{tab:table3} shows that noise degradation impacts the performance of $MFBF_0$ more severely than the pre-trained model representations. We observe that for valence, lexical embeddings are relatively robust to noise than the acoustic embeddings.

Additionally, we explored noise-aware training using the Train+noise set (Table \ref{tab:table1}). We observed improvement across all dimensions for $MFBF_0$ feature, while the noise-aware training had little or no impact on the BERT-based models. BERT embeddings demonstrated robust performance under acoustic distortions. Multi-modal fusion demonstrated overall the best state-of-the-art performance. These findings indicate that the representations learned from the pre-trained models demonstrate robustness against acoustic distortions.

Next, we investigated if representations learned by the multi-modal fusion model can be distilled (using the the procedure outlined in section 5.3) into the $MFBF_0$ and $HuBERT_L$ based models, to improve their robustness. Tables \ref{tab:table4} and \ref{tab:table3} show that knowledge distillation from the multi-modal model helped to improve the robustness of both the models trained with $MFBF_0$ and $HuBERT_L$ features. 
\vspace{-4mm}
\begin{table}[th]
\centering
\caption{Dimensional emotion estimation $CCC$ from uni-modal models and multi-modal fusion model.}
\vspace{1mm}
  \begin{tabular}{lcccccc}
    \toprule
    \multirow{2}{*}{\textbf{System}} &
      \multicolumn{3}{c}{\textbf{Eval1.3}} &
      \multicolumn{3}{c}{\textbf{Eval1.6}} \\
      & {act} & {val} & {dom} & {act} & {val} & {dom} \\
      \midrule
    ${MFBF_0}$ & 0.73 & 0.33 & 0.63 & 0.71 & 0.33 & 0.64  \\
    $BERT_{HuB}$ & 0.33 & 0.56 & 0.29 & 0.30 & 0.55 & 0.27 \\
    $HuBERT_{A}$ & 0.73 & 0.53 & 0.64 & 0.70 & 0.52 & 0.62 \\
    $HuBERT_{L}$ & 0.76 & 0.54 & 0.68 & 0.74 & 0.53 & 0.66 \\
    Multi-modal  & \textbf{0.78} & \textbf{0.68} & \textbf{0.69} & \textbf{0.75} & \textbf{0.66} & \textbf{0.67} \\
    \bottomrule
  \end{tabular}
  \label{tab:table2}
\end{table}
\vspace{-5mm}

\begin{table}[ht]
\centering
\vspace{-2mm}
\caption{Emotion estimation $CCC$ from models trained on Train1.6, Trian+noise and evaluated in noisy conditions.}
\vspace{1mm}
  \begin{tabular}{lcccccc}
    \toprule
    \multirow{2}{*}{\textbf{System}} &
      \multicolumn{3}{c}{Train1.6} &
      \multicolumn{3}{c}{Train+noise} \\
      & {act} & {val} & {dom} & {act} & {val} & {dom} \\
      \midrule
    \midrule
    \textbf{${MFBF_0}$} \\
    $Eval1.6_{25dB}$  & 0.67 & 0.31 & 0.58 & 0.69 & 0.33 & 0.61\\
    $Eval1.6_{15dB}$  & 0.64 & 0.23 & 0.56 & 0.69 & 0.30 & 0.60 \\
    $Eval1.6_{5dB}$   & 0.45 & 0.11 & 0.41 & 0.60 & 0.23 & 0.53 \\
    \midrule
    \textbf{$BERT_{Hub}$} \\
    $Eval1.6_{25dB}$  & 0.31 & 0.55 & 0.27 & 0.29 & 0.55 & 0.26\\
    $Eval1.6_{15dB}$  & 0.31 & 0.54 & 0.27 & 0.29 & 0.54 & 0.26\\
    $Eval1.6_{5dB}$   & 0.29 & 0.51 & 0.26 & 0.28 & 0.51 & 0.26 \\
    \midrule
    \textbf{$HuBERT_A$} \\
    $Eval1.6_{25dB}$  & 0.68 & 0.52 & 0.58 & 0.69 & 0.54 & 0.59 \\
    $Eval1.6_{15dB}$  & 0.65 & 0.51 & 0.57 & 0.68 & 0.53 & 0.58 \\
    $Eval1.6_{5dB}$   & 0.55 & 0.48 & 0.47 & 0.63 & 0.50 & 0.51 \\
    \midrule
    \textbf{$HuBERT_L$} \\
    $Eval1.6_{25dB}$  & 0.72 & 0.51 & 0.61 & 0.71 & 0.51 & 0.61  \\
    $Eval1.6_{15dB}$  & 0.68 & 0.50 & 0.57 & 0.70 & 0.50 & 0.60 \\
    $Eval1.6_{5dB}$   & 0.52 & 0.44 & 0.38 & 0.64 & 0.45 & 0.55 \\
    \midrule
    \textbf{Multi-modal} \\
    $Eval1.6_{25dB}$  & 0.72  & 0.66  & 0.62 & 0.72  & 0.66  & 0.62 \\
    $Eval1.6_{15dB}$  & 0.68  & 0.64  & 0.59 & 0.70  & 0.64  & 0.59 \\
    $Eval1.6_{5dB}$   & 0.53  & 0.59  & 0.51 & 0.57  & 0.57  & 0.43 \\
    \bottomrule
  \end{tabular}
  \vspace{-2mm}
  \label{tab:table3}
\end{table}
\vspace{-1mm}

\section{Conclusions}
\vspace{-2mm}
In this work, we investigated lexical and acoustic representations from pre-trained models and explored their robustness under varying acoustic conditions. We observed that lexical 
\vspace{-3mm}

\begin{table}[H]
\vspace{-3mm}
\centering
\caption{Emotion estimation ($CCC$) from uni-modal models after knowledge distillation from multi-modal model}
\vspace{1mm}
\centering
  \begin{tabular}{lcccccc}
    \toprule
     \multirow{2}{*}{\textbf{System}} &
      \multicolumn{3}{c}{${MFBF_0}$} &
      \multicolumn{3}{c}{$HuBERT_L$} \\
      & {act} & {val} & {dom} & {act} & {val} & {dom} \\
      \midrule
    \midrule
    \textbf{} \\
    $Eval1.6_{25dB}$ & 0.70  & 0.35  & 0.60  & 0.73  & 0.55  & 0.63\\
    $Eval1.6_{15dB}$ & 0.68  & 0.31  & 0.58 & 0.70  & 0.53  & 0.61 \\
    $Eval1.6_{5dB}$ & 0.54  & 0.22  & 0.46 & 0.54  & 0.47  & 0.43 \\
    \bottomrule
  \end{tabular}
  \label{tab:table4}
\end{table}
\vspace{-2mm}
   \noindent representations were helpful for accurate estimation of valence but not for activation and dominance. Low-level acoustic features, such as $MFBF_0$ are found to be susceptible to noise degradation, however, noise-aware training helped to improve its robustness. Models trained with $HuBERT_L$ embeddings were relatively robust compared to $MFBF_0$, especially for valence estimation across SNR levels, however activation and dominance estimation suffered at low SNR levels. Multi-modal fusion helped generate better representations providing the \textit{state-of-the-art} performance that was robust against all SNR conditions. Representations learned from multi-modal fusion and distilling such information to acoustic only systems demonstrated improvement in performance for those systems under noisy conditions. In the future, we plan to investigate how domain mismatch can impact lexical features and valence estimation.



\bibliographystyle{IEEEbib}
\bibliography{custom}

\end{document}